\documentclass[useAMS]{mn2e}
\usepackage{psfig}

\usepackage{times}

\def\lsim{ \lower .75ex\hbox{$\sim$} \llap{\raise .27ex \hbox{$<$}} }
\def\gsim{ \lower .75ex \hbox{$\sim$} \llap{\raise .27ex \hbox{$>$}} }

\title[The Fermi blazars divide ] 
{The Fermi blazars divide}

\author[Ghisellini et al.]
{G. Ghisellini$^1$\thanks{E--mail: gabriele.ghisellini@brera.inaf.it}, 
L. Maraschi$^2$, F. Tavecchio$^1$  \\
$^1$INAF -- Osservatorio Astronomico di Brera, via E. Bianchi 46, I--23807
Merate, Italy\\
$^2$INAF -- Osservatorio Astronomico di Brera, Via Brera, 28, Milano  I--20121
}

\begin{document}



\maketitle

\begin{abstract} 
Flat Spectrum Radio Quasars (FSRQs) and BL Lac objects detected in the first
three months of the {\it Fermi} survey
neatly separate in the $\gamma$--ray spectral index vs 
$\gamma$--ray  luminosity  plane.
BL Lac objects are less luminous and have harder spectra than 
broad line blazars.
We suggest that this division has its origin in the different accretion
regimes of the two classes of objects.
Using the $\gamma$--ray luminosity as a proxy for the observed bolometric one 
we show that the boundary between the two subclasses of blazars can be
associated with the threshold between the regimes of optically thick accretion disks 
and of  radiatively inefficient accretion flows, which lies at an accretion rate
of the order of $10^{-2}$ the Eddington rate.
The spectral separation in hard (BL Lacs) and soft (FSRQs) objects
can then result from the different radiative cooling suffered by the
relativistic electrons in jets propagating in different ambients.
We argue that the bulk of the most luminous blazars already detected by {\it Fermi} 
should be characterised by large black hole masses, around $10^9$ solar masses,
and predict that lowering the $\gamma$--ray flux threshold
the region of the $\alpha _{\gamma}$--$L_{\gamma}$ plane corresponding to steep 
spectral indices and lower luminosities will be progressively populated by FSRQs 
with lower mass black holes, while the region of hard spectra and large 
luminosities will remain forbidden. 
\end{abstract}

\section{Introduction}

The  Large Area Telescope (LAT) on board the {\it Fermi Gamma Ray Space 
Telescope (Fermi)} revealed more than one hundred blazars with a 
significance larger than 10$\sigma$ in the first three months of operation 
(Abdo et al. 2009a, hereafter A09).
Of these, 57 are classified as flat spectrum radio quasars (FSRQs), 
42 as BL Lac objects, while for 5 sources the classification is uncertain. 
Including  2 radio--galaxies the total number of extragalactic sources amounts to 106.
Redshifts are known for all FSRQs, for 30 BL Lacs, for 1 source
of uncertain classification and for the two radio--galaxies,
for a total of 90 objects.
Particularly significant is the large number of BL Lac objects
present in the {\it Fermi} sample, which allows for the first
time an objective comparison of their $\gamma$--ray properties
(spectral shape and luminosity) with those of FSRQs.

A direct result of the first {\it Fermi} blazar sample, discussed by
A09, is the finding of correlations between their $\gamma$--ray and radio 
fluxes/luminosities 
and between the photon spectral index $\Gamma_\gamma$ in the {\it Fermi} 
energy band and their radio luminosity.
In the latter plane, as noted by A09, blazars seem to define a trend, 
of increasing $\Gamma_\gamma$ (softness) for increasing radio luminosity,
with BL Lac objects falling at the lower end of this trend.
However within each of the two separate populations these correlations are 
not highly significant. 

In this letter we consider BL Lacs and FSRQs as part of a unified blazar
population,
at variance with the approach of A09. 
Using the $\gamma$--ray rather than the radio luminosity
(the former being more indicative of the bolometric power),
we discuss the spectral index/$\gamma$--ray luminosity
plane where BL Lacs and FSRQs occupy different regions. 
We then propose a simple scenario to explain why there is a 
rather well defined boundary between BL Lacs and FSRQs.
In our scheme all blazars are essentially similar and their different observed 
properties result from the different accretion rates feeding their central engines.
We show that the observed $\gamma$--ray spectral index dependence on 
$\gamma$--ray (as well as  radio) luminosity can be naturally explained 
in the above context.
Finally we discuss some implications and predictions 
of the proposed scenario.

We adopt a cosmology with $h=\Omega_\Lambda=0.7$ and $\Omega_{\rm M}=0.3$.

\begin{figure*}
\vskip -1 cm
\psfig{figure=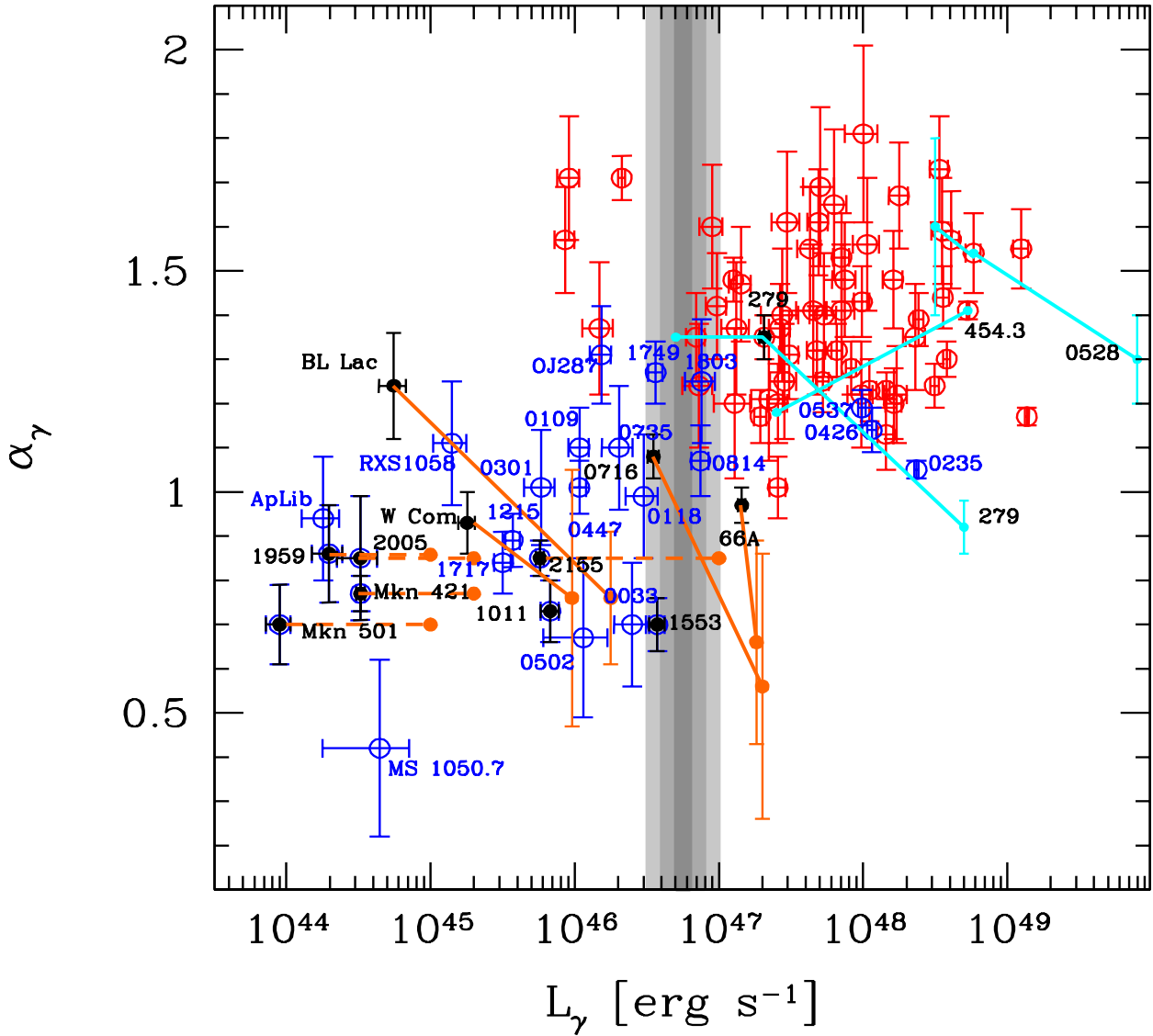,width=18.5cm,height=17.5 cm}
\vskip -1. cm
\caption{
Energy spectral index vs $\gamma$--ray luminosity
for all blazars in the list of Abdo et al. (2009a).
Blue and red points are BL Lacs and FSRQs, respectively.
Black symbols correspond to sources already detected in the TeV band
(they are all BL Lacs but 3C 279).
Labels identify all BL Lac objects, 3C 279 and other 2 FSRQs. 
For several BL Lac objects and for the 3 labelled FSRQs
we show the observed range of $\gamma$--ray luminosity
and spectral index, using past EGRET or {\it AGILE} observations.
This is indicated by a solid line
segment (dashed when the spectral index is unknown, and we chose
the same index as observed by {\it Fermi}). For 0716+714,
the high luminosity point corresponds to {\it AGILE} observations
(Chen et al. 2008) with a large error on the spectral index
(shown by the large error bar).
The grey stripes at about $L_\gamma=10^{47}$ erg s$^{-1}$ 
mark the divide between BL Lac objects and FSRQs.
}
\label{alphal}
\end{figure*}


\section{The gamma--ray slope vs luminosity plane}

We derived the $\gamma$--ray luminosities for all the blazars with known 
redshift detected by {\it Fermi} in the three months survey 
(excluding radio galaxies) 
using the fluxes named $F_{100}$ in 
Table 3 of A09, characterising their ``average" observed
flux  (between August 4 and October 30 2008).
The luminosities were computed according to:
\begin{equation}
L_{\gamma}\, =\, 4\pi d_{\rm L}^2 {S_\gamma(\nu_1, \nu_2)
\over (1+z)^{1-\alpha_\gamma}} 
\label{lg}
\end{equation}
where $d_{\rm L}$ is the luminosity distance, $\alpha_\gamma=\Gamma_\gamma-1$,
$S(\nu_1,\nu_2)$ is the $\gamma$--ray energy flux
between the frequencies $\nu_1$ and $\nu_2$, 
calculated from the photon flux 
$F_\gamma(E>100\, {\rm MeV})$ [ph cm$^{-2}$ s$^{-1}$] as:
\begin{eqnarray}
S_\gamma(\nu_1,\nu_2)\, &=&\, {\alpha_\gamma h\nu_1 
F_\gamma \over 1-\alpha_\gamma} 
\left[ \left( {\nu_2\over \nu_1}\right)^{1-\alpha_\gamma} -1\right]
\quad \alpha_\gamma\ne 1
\nonumber \\
S_\gamma(\nu_1,\nu_2)\, &=&\, h\nu_1 F_\gamma \ln(\nu_2/\nu_1) 
\qquad\qquad \qquad\,  \alpha_\gamma= 1
\label{sgamma}
\end{eqnarray}
Setting $\nu_1=2.42\times 10^{22}$ Hz (0.1 GeV) and
$\nu_2=2.42\times 10^{24}$ Hz (10 GeV), and $F_\gamma=10^{-8} 
F_{\gamma, -8}$ ph cm$^{-2}$ s$^{-1}$,
we have (in units of erg cm$^{-2}$ s$^{-1}$):
\begin{eqnarray}
S_\gamma(0.1,10) &=& 1.6\times 10^{-12}\, {\alpha_\gamma F_{\gamma, -8} 
\over 1-\alpha_\gamma} 
\left[ 100^{1-\alpha_\gamma} -1\right] 
\quad \alpha_\gamma\ne 1
\nonumber \\
S_\gamma(0.1,10) &=& 7.38\times 10^{-12} F_{\gamma, -8}  
\qquad\qquad \qquad\,  \alpha_\gamma= 1
\label{sgamma2}
\end{eqnarray}

Fig. \ref{alphal} shows the energy spectral index $\alpha_\gamma$ given by A09
as a function of $L_\gamma$ between 0.1 and 10 GeV
calculated according to Eqs. \ref{lg}--\ref{sgamma2}.
Blue and red symbols correspond to BL Lac objects and FSRQs, 
respectively, while black symbols indicate sources also detected in 
the TeV band\footnote{We have included 0716+714 as a TeV 
detected BL Lac (Teshima  2008), although
it is not noted as such in A09. Therefore the total number
of {\it Fermi} bright blazar that have been also detected
in the TeV band is 12.
}.
The latter are all BL Lac objects except for 3C 279, which is an 
intermediate object with Ly${\alpha}$ emerging in low states 
(Koratkar, Pian \& Urry 1998; Pian et al. 1999).

Blazars are highly variable sources, especially at high energies.
To illustrate the possible range of variability, for a few sources
we connect with a line segment
the spectral index and $\gamma$--ray luminosity observed by {\it Fermi} 
to the values observed in the past by the EGRET instrument onboard the 
{\it Compton Gamma Ray Observatory} satellite (Nandikotkur et al. 2007), 
or observed by the {\it AGILE} satellite.
Note that variable sources often ``move" orthogonally to the 
correlation defined by the ensemble of sources, i.e.
they become harder when brighter (with the exception of 3C 454.3).

Moreover, for some of the sources detected in the TeV range,
we can infer the level of $\gamma$--ray 
activity during high states of TeV emission (assuming that the flux in 
the {\it Fermi} band would have been about the same as the TeV flux).
For these sources we indicate the
range of luminosity variability by dotted line segments,
maintaining  the spectral index measured by {\it Fermi} for both states. 

As can be seen, the high and the low $\gamma$--ray states of single
sources can be dramatically different. 
This implies that the distribution in luminosity 
within each blazar class is largely affected by
the variability of the sources.


The $\alpha_\gamma$--$L_\gamma$ plane (Fig. \ref{alphal}) strongly suggests
a separation  of the two groups in regions defined by $\alpha_\gamma>1.2$,
$L_\gamma>10^{47}$ erg s$^{-1}$ for FSRQ and $\alpha_\gamma<1.2$, 
$L_\gamma<10^{47}$ erg s$^{-1}$ for BL Lacs. 
At the same time there is continuity between the two 
blazar subclasses in that their properties overlap at intermediate
values of both spectral index and  $\gamma$--ray luminosity.  

The separation in spectral indices, already evident in Fig. 9 of A09, 
occurs at $\alpha_\gamma\simeq 1.2$
Only 5 FSRQs have spectra flatter than 1.2, and only 5 BL Lacs 
have spectra steeper than 1.2.  
The LAT $\gamma$--ray data have been systematically analysed by A09 
fitting a single power law model to the spectral data of each source.
In many cases the intrinsic spectral shape could be more complex 
(i.e. a curved spectrum). 
Fig. 10 of A09 shows as illustrative examples the $\gamma$--ray SEDs of 3C 454.3 
(energy index $\alpha_\gamma=1.41\pm0.02$), A0 0235+164 ($\alpha_\gamma=1.05\pm0.02$) 
and Mkn 501 ($\alpha_\gamma=0.70\pm0.09$).
In the first two cases the data are not well represented
by the single power--law model, showing significant curvature and indicating
SED peaks below 100 MeV for 3C 454.3 and between 1 and 10 GeV for A0 0235+164, 
while Mkn 501 shows a very hard $\gamma$--ray SED implying that
the SED peak should be above 100 GeV.
It is therefore reasonable to assume that the spectral index measured by 
{\it Fermi} is a proxy for the peak energy of the $\gamma$--ray SED, 
at least in a statistical sense.
Thus the $\gamma$--ray SEDs of most FSRQs, with $\alpha_\gamma >1.2$,
should peak at or below the low energy range of the {\it Fermi} energy
window, at lower peak energies than those characterising most of  
BL Lacs objects.
As expected, BL Lacs detected in the TeV band tend to be harder than 
other BL Lacs (with the exception of BL Lac itself), indicating a 
high peak energy, in turn necessary to detect them in the TeV band.

The separation in $\gamma$--ray luminosity between BL Lac objects and FSRQs
is also striking.
There are only 3 BL Lac objects (PKS 0537--441 with $z=0.892$, 
AO 0235+164 with $z=0.94$ 
and PKS 0426--380 with $z=1.12$) with $L_\gamma>10^{47}$ erg s$^{-1}$.
The redshifts of these 3 sources are among the highest of the BL Lac 
sub--sample (see Fig. 16 of A09).
Moreover, {\it all} three sources 
{\it do have broad lines}, visible in their low emission states
(see Sbarufatti et al. 2005 for PKS 0426--380;
Pian et al. 2002 for PKS 0537--441; Raiteri et al. 2007 for AO 0235+164).

In principle, the two blazar classes could be
separated because of a different average Doppler boosting,
but all indications so far 
for the $\gamma$--ray emitting BL Lacs and FSRQs suggest instead
similar values of the Doppler boosting,
with the exception of TeV emitting BL Lacs, requiring
a stronger boost (and thus a smaller viewing angle).
These inferences come from superluminal motion
(Jorstad et al. 2001;
Kellerman et al. 2004),
direct modelling of their SED 
(e.g. Konopelko et al. 2003; Finke, Dermer \& B\"ottcher 2008)
and very rapid TeV variability (e.g. Begelman, Fabian \& Rees 2008;
Ghisellini \& Tavecchio 2008b).
Thus the idea that the luminosity sequence  
of BL Lacs and FSRQs is due to larger viewing angles for BL Lacs 
is disfavoured.

\subsection{The blazar sequence in $\gamma$--rays}

The SEDs of blazars show two main components: the first one, peaking between the 
far infrared  and the X--ray band, is unanimously attributed to 
synchrotron emission; the second, peaking between 10 MeV--100 GeV,
is widely attributed to the Inverse Compton process 
involving the same electrons producing the synchrotron component 
(e.g. Maraschi et al. 1992; Bloom \& Marscher et al. 1996;
Sikora et al. 1994; Dermer \& Schlickeiser 1993).

Fossati et al. (1998) and Ghisellini et al. (1998) proposed
that the SEDs of blazars form a ``spectral sequence"
whereby the peaks of the two components are governed by luminosity, 
and are at higher energies for lower luminosities\footnote{The original
idea considered the luminosity as the main parameter governing the blazar sequence, 
while the refined version (Ghisellini \& Tavecchio 2008a)
considers the luminosity in Eddington units.}.
This was interpreted in terms of a stronger radiative cooling for more
powerful sources, resulting in smaller energies of the electrons emitting at 
the peaks.
As discussed above, a high energy peak above $\sim$10 GeV corresponds to a 
hard spectrum in the {\it Fermi} range (rising, in $\nu F_\nu$), while a 
peak frequency below $\sim$1 GeV corresponds to a soft spectrum 
in the {\it Fermi} range. 
Therefore the trend apparent in the $\alpha_\gamma$--$L_\gamma$ plane 
strongly supports the ``sequence" concept, which was proposed on the 
basis of radio and X--ray selected samples, where only a fraction of 
objects had $\gamma$--ray data.
Fig. \ref{alphal} represents the $\gamma$--ray selected version of the
blazar sequence.

Single objects, instead, often behave in the opposite way (i.e. harder
when brighter). 
This can be explained, if, in single objects, an unchanged cooling 
faces a fast increasing heating (i.e. injection of more energetic electrons),
shifting the typical electron energies to larger values.
In the next section we present a possible interpretation of the separation of 
FSRQs and BL Lacs in the $\alpha_\gamma$--$L_\gamma$ plane.

\section{Interpretation of the BL Lac -- FSRQ divide}

The neat separation between BL Lacs and FSRQs in the 
$\alpha_\gamma$--$L_\gamma$ plane together with their close contiguity 
are really remarkable. A possible interpretation could simply be 
to assume two independent populations with different physical conditions 
as to the origin, propagation and radiative properties of their jets.  
This however would not account for the contiguity which should be
coincidental.
Figs. 11 and 17 of A09 ($\alpha_\gamma$ vs $z$ and $L_\gamma$ vs $z$) 
indicate that the separation and the contiguity behaviour of the two 
blazar sub--classes exist also in these planes.

Building on previous work on the systematic properties of blazars we 
suggest that this division has a physical origin, related to the
different mass accretion rates onto the central engine.
We first summarise some preliminary steps as follows:

\begin{enumerate}


\item
Radiative models allow to estimate the total jet power $P_{\rm j}$ 
carried in the form of bulk and internal
energy of protons, electrons and magnetic field. For powerful FSRQs 
a fraction $\epsilon \sim 0.1$ of $P_{\rm j}$  is radiated
at short distances (i.e. hundreds of Schwarzschild radii)
from the central black hole.
This fraction can be larger for BL Lacs (Celotti \& Ghisellini 2008).

\item
In powerful FSRQs, $P_{\rm j}$ is about one order of magnitude larger than
the {\it observed} accretion disk luminosity $L_{\rm d}$
(Maraschi 2001; Maraschi \& Tavecchio 2003; Celotti \& Ghisellini 2008;
Ghisellini \& Tavecchio 2008a; see also the earlier results of
Rawlings \& Saunders 1991). 
For a ``standard" Shakura \& Sunyaev (1973)
accretion disk, $L_{\rm d} = \eta \dot M c^2$, where $\dot M$ is the accretion
rate, with $\eta \sim 0.1$.
It follows that the jet power $P_{\rm j}$ is of the same order as the
accretion power $\dot M c^2$.
This is also supported theoretically
(at least at the ``dimensional level") since, even if the black hole rotation 
played an essential role as in the Blandford \& Znajek (1977) model, accretion would 
still be needed to provide the magnetic field necessary to extract the 
rotational energy.
We then assume that the (order of magnitude) equality  
$P_{\rm j} \simeq \dot M c^2$ has general validity for strong jets,
being aware of the significant uncertainties associated with it.

\item
Finally, the $\gamma$--ray luminosity  dominates
the electromagnetic output of powerful FSRQs and is at least
a sizeable fraction of it in less powerful BL Lacs.
The observed $L_\gamma \sim  \Gamma^2 \epsilon P_{\rm j}$
(the factor $\Gamma^2$ accounts for light aberration).

\end{enumerate}

Adopting the relations above, it then follows that 
the maximum observed $L_\gamma$ should correspond to the largest black hole 
masses accreting at the maximum, Eddington, rate
\begin{equation}
L_{\rm \gamma, max} \sim \epsilon \Gamma^2 \dot M_{\rm Edd} c^2  
\sim 1.5\times 10^{47} \epsilon \Gamma^2 M_9 \,\,\,
{\rm  erg~~s^{-1}}
\end{equation}
where $M_9$ is the mass of the black hole in units of $10^9$ solar masses.
The $\gamma$--ray luminosity of the most luminous FSRQs (Fig. \ref{alphal})
is of the order of several $\times 10^{48}$ erg s$^{-1}$,
corresponding to $\epsilon \Gamma^2 M_9 \sim 30$.

Fig. \ref{alphal} shows a divide at a $\gamma$--ray luminosity 
roughly a factor 100 below the maximum (grey stripes).
Maintaining the same value of $\epsilon \Gamma^2 M_9 \sim 30$,
we conclude that the jet power, hence the accretion rate,
is a factor $\sim$100 below the maximum. 
Therefore, at the dividing luminosity, the accretion rate is of order:
\begin{equation}
\dot M_{\rm div} \approx { 10^{-2} \dot M_{\rm Edd} }
\end{equation}
Crucial, in this simple derivation, is the assumption that the 
most luminous FSRQs and the most luminous
BL Lacs (excluding the three ``outliers" mentioned
above) have the same factor $\epsilon \Gamma^2 M$, and
in particular the same black hole mass $M$.
This is likely, since we are dealing with the  most 
luminous objects of the two classes (this also applies to the original 
blazar sequence).
Indeed, for TeV BL Lacs, which span the entire region of BL Lacs,  
the available estimates indicate high black hole masses (e.g. Wagner 2008, 
Ghisellini \& Tavecchio 2008a), in some cases exceeding  $10^9  M_{\odot}$.
In the case of FSRQs (more luminous than BL Lacs)
{\it Fermi} could have already started to explore smaller 
black hole masses.
In fact, the object PMN 0948+0022 ($\alpha_\gamma=1.6\pm 0.14$ and
$L_\gamma \sim10^{47}$ erg s$^{-1}$) is a NLSy1, with a black hole mass
around $10^8M_\odot$ (see Abdo et al. 2009b and Zhou et al. 2006). 
This easily accounts for the few FSRQs with $L_\gamma<10^{47}$
erg s$^{-1}$ (i.e. they should have black hole masses smaller
than the rest of FSRQs).

The derived value of $\dot M_{\rm div}$ is naturally interpreted 
as due to the transition between different accretion regimes:
for high accretion rates, in Eddington units,
a standard optically thick 
geometrically thin, radiatively efficient Shakura--Sunyaev (1973) disk
is formed producing a luminous Broad Line Region through which the jet must
propagate suffering strong radiative losses. 
For $\dot M\sim\dot M_{\rm Edd}$ some of the disk luminosity
could be advected into the black hole, resulting in $\eta<0.1$,
but this would not affect the presence of the BLR, nor the
power of the jet.

When $\dot M\ll \dot M_{\rm Edd}$
the accretion flow becomes optically thin and radiatively inefficient
and could be described as an ion supported torus (Rees et
al. 1982) or an advection dominated accretion flow
(ADAF, see e.g. Narayan, Garcia \& McClintock 1997), an adiabatic
inflow--outflow (ADIOS, Blandford \& Begelman 1999) or a
convection dominated flow (CDAF, Narayan, Igumenshchev \& Abramowicz 2000).
In the latter case despite the uncertainties in the dynamics of the flow
the surrounding region will certainly be deprived of photons with respect 
to the high accretion rate regime, thus justifying the absence of broad 
emission lines and the reduced radiative losses for the propagating jet.

Since FSRQs are associated with radio sources with FR II morphology
while BL Lac objects are usually found in FR Is, the same critical rate should 
separate the two parent populations. 
And indeed the finding by Ledlow \& Owen 
(1996) that the radio luminosity threshold between the two classes 
depends on the optical magnitude of the bulge of the host galaxy 
(i.e. on the mass of the central black hole), was interpreted by 
Ghisellini \& Celotti (2001) in terms of a critical accretion rate,
coincident in value with what found here 
(see also more recent work by Xu, Cao \& Wu 2009).

\section{Discussion and conclusions}

The spectral shape/$\gamma$--ray luminosity plane revealed
a clear separation between FSRQs and BL Lac objects.
Moreover, there is a well defined trend, with BL Lac objects
being harder and less luminous than FSRQs.
At the first order, this strongly confirms the idea that
blazars obey a spectral sequence, controlled by the bolometric
luminosity, of which the $\gamma$--ray one is good proxy.

The other striking feature is the little overlap, in 
$\gamma$--ray luminosity, of the two classes of blazars.
We suggest that this has a physical origin,
and have interpreted it in terms of a critical accretion rate 
$\dot M_{\rm div}$.
Below  $\dot M_{\rm div}$ accretion becomes radiatively inefficient,
the corresponding ionising flux becomes weaker, making 
weaker (or absent) broad emission lines.
These are BL Lac objects. 
Their jets thus propagate in a medium starved of external radiation
(weak disk, weak lines), and this makes the emitting electrons accelerated 
in the jet to cool less, and to reach very high energies.
Their emitted high energy spectrum, produced 
mainly by the synchrotron self--Compton mechanism, is less luminous and 
harder than in FSRQs.
The latter sources in fact have a radiatively efficient accretion
disk and a corresponding ``standard" broad line region.
If the jet dissipates most of its power within the broad line region, then
the emitting electrons will efficiently cool 
(mostly by external Compton), will reach only moderate energies,
and will produce a high energy peak below 100 MeV.
This scenario is in perfect agreement with the analogous division
proposed for radio--galaxies, that in turn are though to be
the progenitors of BL Lacs (FR I) and FSRQs (FR II).

According to this idea the power of the jets in all blazars is
proportional to the accretion rate, that can change in 
a continuous way from one objects to the other (and even
in single objects). 
The observed discontinuity between BL Lacs and FSRQs is
the result of a discontinuity in the accretion regime
occurring within a rather narrow range of $\dot M$
(from standard to radiatively inefficient).
Thus, despite the different ``look" of FSRQs and BL Lacs,
they all belong to a sequence of jet powers. From 
this point of view, it is therefore reasonable to 
treat them as a single class.

Evidences that $P_{\rm j}$ is of the same order of $\dot Mc^2$ can be traced
back to Rawlings \& Saunders (1991), and since then confirmed 
by several works (e.g. Celotti et al. 1997; 
Cavaliere \& D'Elia 2002; D'Elia, Padovani \& Landt 2003;
Maraschi \& Tavecchio 2003; Sambruna et al. 2006;
Allen et al. 2006;
Celotti \& Ghisellini 2008; Kataoka et al. 2008).
Also, the discovery that large scale jets of powerful blazars are 
strong X--ray emitters,
detectable and resolvable by the {\it Chandra} satellite, offered a novel way
to measure $P_{\rm j}$, confirming that in these sources the jet power
is not only of the same order than $\dot M c^2$, but it 
is also constant all along the jet (Tavecchio et al. 2004; Cheung et al. 2006).

Also the idea that the differences between powerful FSRQs (and
their FR II radio--galaxy counterpart) and BL Lacs (and FR I radio--galaxies)
is due to a difference in the accretion regime of their disks has been
already put forward (Ghisellini \& Celotti 2001; Jester 2005) and 
used to explain the different cosmic evolution of these two
classes of blazars (see e.g. B\"ottcher \& Dermer 2002;
Cavaliere \& D'Elia 2002).
Interestingly, also in radio--quiet AGNs there is evidence 
that the relative strength of the broad emission lines is a function
of the accretion disk luminosity in Eddington units,
with the possibility of a very weak BLR for accretion rates
below $10^{-2} L_{\rm Edd}$ (for review, see Ho 2008).
What is new, in this paper, is the interpretation, in terms
of accretion rates, of the intriguing separation of 
FSRQs and BL Lacs detected by {\it Fermi} in the 
$\alpha_\gamma$--$L_\gamma$ plane. 
A (likely) cosmic evolution of $\dot M$, while not affecting our results,
would account for the existing hints of a negative and positive
cosmic evolution of BL Lacs and FSRQs, respectively (see, e.g. A09).

There are a few checks to put our scenario on test.
Perhaps the simplest is to have clues on the black hole
masses of the most luminous BL Lacs and FSRQs.
We predict them to be of the same order.
A related check is to directly find the accretion disk luminosity
for the most luminous FSRQs. 
If also the black hole mass is known or derived, then we can test if their 
disk are indeed emitting close to the Eddington limit, as we have assumed.

Although exceptions exist for any rule, it should be 
interesting to study in more detail the three 
BL Lacs that are ``outliers", lying in the FSRQ--region.
Since they do have broad lines, we propose that they are normal FSRQs
whose non--thermal continuum is so enhanced by beaming that
it hides the broad lines.
If this is the case, the FSRQs/BL Lacs division, now
in terms of the line equivalent width, could be substituted
by a more physical division in terms of broad line luminosity.

In just 3 months of all sky survey (and selecting only the brightest ones), 
{\it Fermi} has already doubled the number of known $\gamma$--ray 
emitting blazars.
Soon we will have to deal with thousands of them.
Since blazars are very variable, perhaps spending a limited amount 
of time in high state, continuing to patrol the all sky will surely 
reveal exceptional states of sources otherwise unnoticeable.
On the other hand, the increase in sensitivity due to the increased
exposure time will allow to find many less powerful jets and less luminous
$\gamma$--ray luminosities. 
The divide will then be not as clear as it is now, since many FSRQs
(with black hole masses below $10^9M_\odot$)
will occupy the region below $L_\gamma=10^{47}$ erg s$^{-1}$.
On the other hand, the prediction is that no or very few BL Lacs will
be found above this limit.
These improved samples will surely be useful to confirm, or reject,
or constrain our scenario.

\section*{Acknowledgments}
This work was partly financially supported by a 2007 COFIN--MIUR grant.
We thank the referee for useful comments.

\end{document}